# Interfacing of High Temperature Z-meter Setup Using Python

Ashutosh Patel[*,1], Shashank Sisodia[1,2] and Sudhir K. Pandey[1]

[1]*School of Engineering, Indian Institute of Technology Mandi, Kamand 175005, Himachal Pradesh, India*
[*]*Email:ashutosh_patel@students.iitmandi.ac.in*

**Abstract.** In this work, we interface high temperature Z-meter setup to automize the whole measurement process. A program is built on open source programming language 'Python' which convert the manual measurement process into fully automated process without any cost addition. Using this program, simultaneous measurement of Seebeck coefficient ($\alpha$), thermal conductivity ($\kappa$) and electrical resistivity ($\rho$), are performed and using all three, figure-of-merit (ZT) is calculated. Developed program is verified by performing measurement over p-type $Bi_{0.36}Sb_{1.45}Te_3$ sample and the data obtained are found to be in good agreement with the reported data.

## INTRODUCTION

ZT is an important parameter which represents the applicability of the thermo-electric (TE) materials. It depends on the $\alpha$, $\kappa$ and $\rho$, defined as $ZT=\alpha^2 T/\kappa\rho$. Where T is the mean temperature across TE sample.[1] Finding the value of ZT requires the measurement of $\alpha$, $\kappa$ and $\rho$. All three parameters are temperature dependent. Continuous measurements of all three parameters are required for wide temperature range, which take long time. Manual operation of measurement requires continuous monitoring. A similar instrument to measure ZT is reported in Ref. [2], where LabVIEW is used to automize the measurement process. LabVIEW, Visual Basic, etc are commercial softwares and available at high cost.

In this work, we have developed a program based on open source programming language 'Python' to interface Z-meter setup with power supply unit and digital multimeter. This program makes the measurement process fully automated. At the start of the program all control parameters are fed and further no user interaction is required throughout the measurement. PyVISA, NumPy, matplotlib.pyplot libraries are used in program for hardware communication, mathematical operations and live-plotting of data, respectively. This program measures $\alpha$, $\kappa$ and $\rho$, simultaneously and using them ZT is calculated. Use of open source program reduced the setup cost largely. P-type $Bi_{0.36}Sb_{1.45}Te_3$ sample is used to verify the program and collected data are found to be in good agreement with the reported data.

## INTERFACING OF Z METER

High temperature Z meter setup requires power supply to heat the sample and 4-probe current for resistivity measurement for which dual channel Keithley 2604B sourcemeter (SMU) is used. Measurement of all raw data is performed by using Keithley 2002 digital multimeter (DMM). This multimeter contains a 10 channel scanner card, which enable this to measure the number of data. To communicate both hardware with computer, IEEE-488B GPIB interface along with GPIB-USB converter is used.

A program is built on open source programming language "Python". PyVISA, NumPy, matplotlib libraries are used for hardware communication, mathematical operations and real-time plotting of $\alpha$, $\kappa$, $\rho$, and ZT. First all control parameters like sample dimension, measurement temperature limit, step power etc are defined. Based on these parameters, SMU supplies current to the heater and wait to reach steady state. It is ensured by monitoring hot side temperature ($T_h$). An increase in $T_h$ with time is obtained. The screenshot of the program used for steady state is shown in Fig. 1. $T_h$ is monitored continuously with an interval of 5 second and its rate of change is obtained which is shown in Fig. 2. From this figure, it is clear

```
th1 =float(machine1.ask(":READ?\r"))
#print temperature1
time.sleep(5)
th2 =float(machine1.ask(":READ?\r"))
#print temperature2
a = float(th2 - th1)            # increase in Th
if a > 0 :
    a = a
else:
    a = -a
print ("increase in Th %f" %a )  # temperature difference
#print cnt
axis.plot(count,temperature1, 'o') # plot Th
plt.pause(0.2)
axis0.plot(count,a, 'o')         # plot increase in Th
count = count + 1
if(temperature2>1000):

    plt.pause(1)

if (a > 0.006):
    cnt = 0
else:
    if(a<0.006):
        cnt = cnt + 1
if(cnt==5):
```
**FIGURE 1.** Program to ensure steady state.

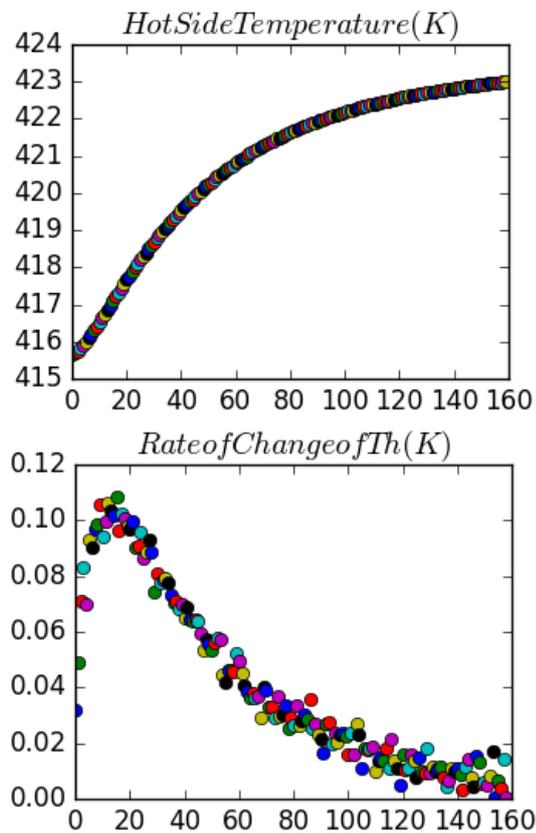

**FIGURE 2.** Change in $T_h$ and its rate of change.

that the rate of change of $T_h$ decreases with time. A very small rate of change (0.006 K per measurement) is defined as steady state criteria. To ensure steady state more accurately, program doesn't allow to measure raw data until the condition is satisfied successively five times. Once steady state is ensured, the measurement of raw data starts. The program used for it is shown in Fig. 3. First temperature at both ends are measured. Based vacuum chamber temperature and $T_h$ heat loss is calculated using interpolation. Thermolectric voltage is measured at the negative ends of both thermocouples. After this, all required parameters are calculated. Real-time plotting of all these parameters are implemented by using matplotlib.pyplot. Along with all raw data, these parameters are exported to a .csv file.

*Measurement Setup*

The measurement setup contains a sample holder in which sample is to be fixed. There are two thermocouples to measure the temperature across the sample end. Thermoelectric voltage of sample is obtained by measuring voltage across negative legs of thermocouples. This sample holder assembly is placed inside the vacuum chamber. And electrical connector is available to make electrical connections. Diffusion pump with rotary backing is used to create a vacuum of level $10^{-5}$ mbar. During whole measurement process the room temperature is mentioned at 300 K to measure heat loss accurately.

```
machine1.write(":ROUT:CLOS (@1)")
Th1 =float(machine1.ask(":READ?\r"))   # hot side temperature
machine1.write(":ROUT:CLOS (@2)")
Tc2 =float(machine1.ask(":READ?\r"))   # cold side temperature
smu_Power =float( machine2.ask("print (smub.measure.p())") )
smu_Current =float( machine2.ask("print (smub.measure.i())"))
machine1.write((":SENS:FUNC 'VOLT:DC'"))
machine1.write(":ROUT:CLOS (@7)")
dmm_Voltage =float( machine1.ask(":READ?\r"))
time.sleep(1)
machine1.write(":ROUT:CLOS (@3)")
machine1.ask(":READ?\r")
ch3 =float( machine1.ask(":READ?\r"))
v3 = ch3*(10**(6))
machine1.write(":ROUT:CLOS (@4)")
machine1.ask(":READ?\r")
ch4 = float(machine1.ask(":READ?\r"))
v4 = ch4*(10**(6))
machine1.write(":ROUT:CLOS (@3)")
machine2.write("smua.source.leveli = {0}".format(ir))
v1 = float(machine1.ask (":READ?\r"))
machine2.write("smua.source.leveli = {0}".format(-ir))
v2 = float(machine1.ask (":READ?\r"))
```
**FIGURE 3.** Program to aquire raw data.

## RESULTS AND DISCUSSIONS

This program developed to automate the measurement process of Z meter setup is validated by using p-type $Bi_{0.36}Sb_{1.45}Te_3$ sample of 1.4mm*1.4mm cross section and 1.7 mm of thickness. This sample is extracted from a commercially available thermoelectric generator and composition is obtained by performing EDX analysis. The measured value of α, κ, ρ, and ZT are shown in Fig. 4. Measurement performed from $T_h$=320 K to $T_h$=560 K. Temperature difference across sample varies almost linearly from 12 K to 90 K where $T_h$ changes from 320 K to 560 K. Our measured data are compared from reported data.[3] At T=315 K, the value of α and κ match closely with the reported data, but the measured value of ρ is having a deviation of 0.24 mΩ.cm. As the temperature increases deviation in α and κ increases while deviation in ρ is decreased. At T=500 K, the deviation in α and κ are 25 μV/K and 0.8 W/m.K, respectively. At this temperature measured value of ρ matches closely with reported data. As ZT is obtained by combining all three α, κ and ρ, At T=315 K, deviation of 0.2 is observed in the measured value compared with the reported value. This deviation decreases with increase in temperature and at T=500 K it is about 0.1.

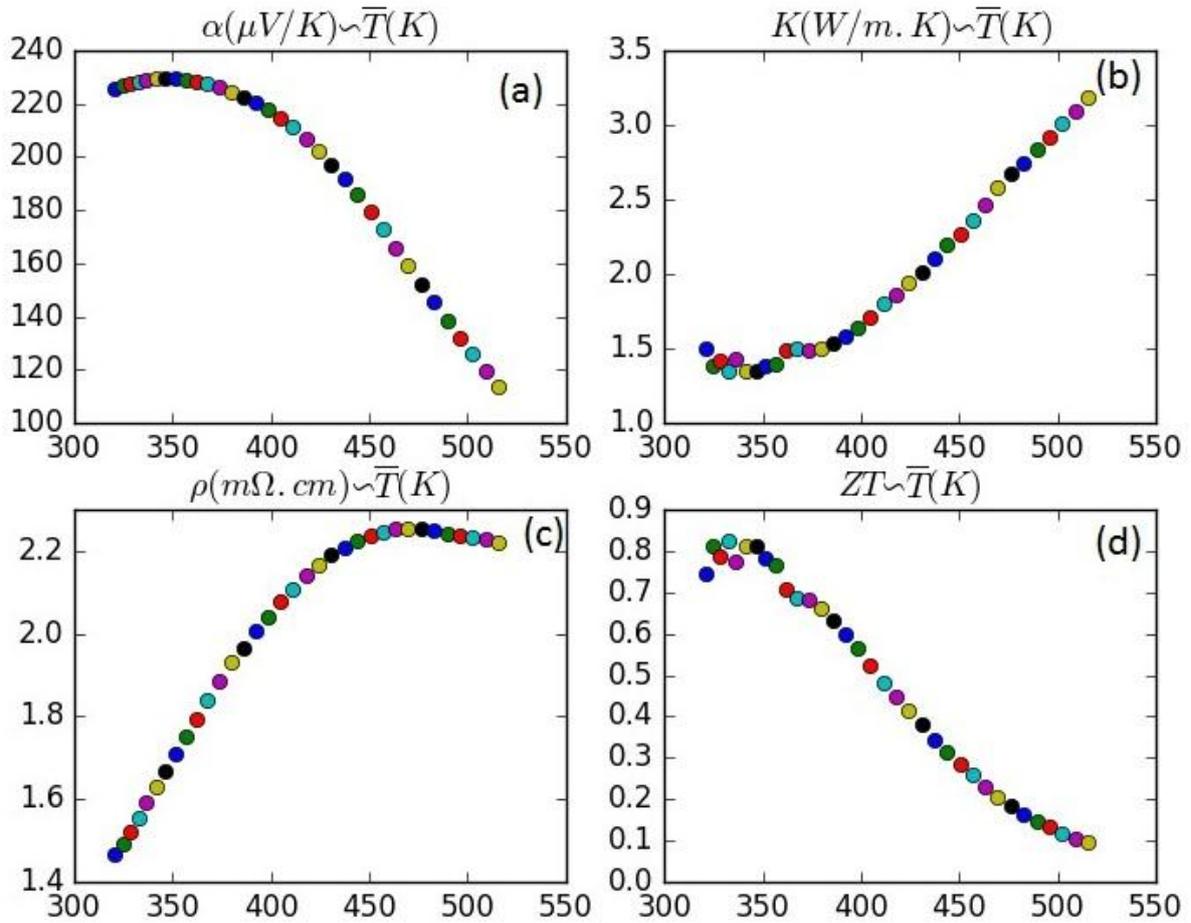

**FIGURE 4.** (a) Seebeck coefficient, (b) Thermal conductivity, (c) Electrical resistivity, and (d) ZT, at different temperature of $Bi_{0.36}Sb_{1.45}Te_3$ sample.

## CONCLUSIONS

We have interfaced a Z-meter setup using open source programming language 'Python'. A manual measurement process converted into a fully automated process, without any cost addition, due to the use of open source platform. This program is validated by performing measurement of $Bi_{0.36}Sb_{1.45}Te_3$ sample and measured data were found in good agreement with the reported data.

## ACKNOWLEDGEMENTS

The author Shashank Sisodia is thankful to IIT Mandi for the internship and financial support.

[2] *Present address- Mechanical and Automation Engineering, Maharaja Agrasen Institute of Technology, Delhi 110086, India*